# The Orion OB1 association

## II. The Orion-Eridanus Bubble

A.G.A. Brown[1], D. Hartmann[1,*] and W.B. Burton[1]

[1] Sterrewacht Leiden, P.O. Box 9513, 2300 RA, Leiden, The Netherlands

**Abstract.** Observations of the interstellar medium in the vicinity of the Orion OB1 association show a cavity filled with hot ionized gas, surrounded by an expanding shell of neutral hydrogen (the Orion-Eridanus Bubble). In this paper we examine this cavity and the surrounding bubble with the aid of data from the Leiden/Dwingeloo H I survey. We present neutral-hydrogen maps for the Orion-Eridanus region which allow identification of the H I filaments and arcs delineating the Bubble and derivation of its expansion velocity. The X-ray emission from the Orion-Eridanus region is enhanced with respect to the Galactic background emission. Comparison with the H I data shows a detailed anti-correlation of the X-ray enhancement with kinematically-narrow features. This confirms the association of the X-ray enhancement with the cavity in H I. Comparison of the derived column densities in H I with the IRAS 100 $\mu$m intensities shows that the H I shell emission is optically thin. This justifies a derivation of the mass of the H I shell by direct conversion of the observed H I emission to column densities. Models of wind-blown bubbles are considered, to investigate if the cavity was formed by the stellar winds and supernovae from Orion OB1. Using a model that takes the density stratification of the Galactic H I layer into account, we show that the stellar winds and supernovae from stars in Orion OB1 can account for the size as well as for the expansion velocity of the H I shell. However, density inhomogeneities in the ambient interstellar medium cause significant discrepancies between our model and the observed shell.

*Send offprint requests to*: A.G.A Brown

* Present address: Harvard-Smithsonian Center for Astrophysics, Mail Stop 47, 60 Garden Street, Cambridge, MA 02138, USA

## 1. Introduction

OB associations play an important role in the propagation of star formation and in the evolution of the interstellar medium, in general marshalling various aspects of the structure and the kinematics of the medium. A prototypical example of the marshalling of the ISM by an association can be found around Orion OB1. This stellar association consists of four Subgroups (1a–1d, see Blaauw 1964) and is accompanied by several large-scale features in the interstellar medium. These features include the Orion Molecular Cloud (OMC) complex (see e.g. Kutner et al. 1977 and Maddalena et al. 1986); Barnard's Loop (see e.g. Goudis 1982); H$\alpha$ emission structures which extend all the way to the Eridanus constellation, some 50° below the Galactic plane (Sivan 1974; Reynolds & Ogden 1979, henceforth RO); and a cavity in the neutral hydrogen distribution surrounded by H I arcs showing expansion motions (Heiles 1976; Green 1991; Green & Padman 1993). The diffuse Galactic H$\alpha$ background intensity is enhanced in this area (e.g., Reynolds et al. 1974), and the same is true for the X-ray intensity (e.g., Burrows et al. 1993, henceforth BSNGG). Cowie et al. (1979) measured ultraviolet absorption lines of various ionization stages of C, N, Si, and S for stars in or near Orion OB1 and detected a high-velocity, low-ionization shell surrounding all of the Orion area ("Orion's Cloak"), as well as a dense, clumpy, low-ionization shell expanding at smaller velocity. The cavity in the H I emission also shows up in the IRAS sky flux maps of the Orion-Eridanus area shown by BSNGG.

The various observed features were interpreted by RO as being consistent with a cavity of warm ionized gas surrounded by an expanding shell of H I. The ionization of the cavity would be maintained by the UV radiation from the early-type stars in Orion OB1. They explained the existence of the shell as due to a series of supernova explosions about $2 \times 10^6$ years ago. Cowie et al. (1979) interpreted the high-velocity shell in terms of the radiative shock of a supernova which occurred some $3 \times 10^5$ years ago, and the high H I column densities in terms of the remnants of older supernova events which occurred 2–4 $\times 10^6$ years ago.

More recently, BSNGG presented a model in which part of the observed X-ray enhancement is due to hot gas filling a cavity created by the stellar winds from early type stars in Orion OB1. Both RO and BSNGG present schematic views of the Orion-Eridanus Bubble in which it expands towards the Sun. In the model of BSNGG the Orion-Eridanus Bubble interacts with the Local Hot Bubble in which the Sun is evidently immersed.

In this paper we present a more detailed look at the neutral hydrogen content of the Orion-Eridanus region, using the Leiden/Dwingeloo H I survey. This H I survey has better spatial- and velocity-resolution than the surveys used previously for studies of the Orion-Eridanus Bubble. To model the size and expansion velocity of the Bubble we use detailed information on the stellar content of Orion OB1 as described by Brown et al. (1994, henceforth Paper I). We employ models which take into account the density stratification of the Galactic H I layer.

This Paper is organized as follows. In Sect. 2 we give a short description of the Leiden/Dwingeloo H I survey and the characteristics of the data. In Sect. 3 we present the H I data for the region of interest. A comparison with data at other wavelengths is made in Sect. 4, in particular with the BSNGG X-ray material. In Sect. 5 we derive column densities of the H I filaments and the mass of the neutral hydrogen shell and discuss the energetics of the Bubble and its relation to the stars in Orion OB1. Section 6 contains the conclusions.

## 2. H I survey data in the region

The H I data used in this paper are from the recently completed Galactic survey made with the 25 m Dwingeloo telescope[1] (Hartmann 1994 and Hartmann & Burton 1995). The entire sky north of declination $-30°$ was mapped in the 21 cm emission line, using a helium-cooled FET amplifier receiver with a system temperature of about 35 K, together with the 1024-channel prototype of the Dwingeloo Auto-correlation Spectrometer (DAS). The telescope beam width is $36'$; the observations were sampled on a $0°\!.5$ lattice typically with 180 s integration time. The data were obtained in total-power mode. Frequency-switched bandpass calibrations were frequently observed at one full bandpass (5 MHz) higher frequency. After data reduction, the velocity coverage of the survey is $-450 \leq v_{\rm LSR} \leq +400\,{\rm km\,s^{-1}}$ at $1.03\,{\rm km\,s^{-1}}$ resolution. (All velocities quoted here are expressed relative to the Local Standard of Rest.) The mean sensitivity of the data is about 0.07 K.

The data were corrected for stray-radiation contamination using the method of Kalberla (1978; see also Kalberla et al. 1980). His algorithm, originally developed for the Effelsberg 100 m telescope antenna pattern and using the Berkeley H I survey (Weaver & Williams 1973; Heiles & Habing 1974) as input, was adapted to calculate the stray radiation contribution for all spectra in the current survey. This led to convolving the measured Leiden/Dwingeloo input sky with the 25 m's antenna pattern. Details of this procedure are given in Hartmann (1994) and Hartmann & Burton (1995). The calculated stray radiation was subtracted from the data and a third-order polynomial baseline was removed using an unbiased automated algorithm. A low-amplitude (typically $\sim 0.008$ K) standing wave was subtracted, and sharp interference spikes were detected, modeled, and removed.

The reduced data were re-sampled onto a common grid ($0°\!.5 \times 0°\!.5$) to create a homogeneous data cube of dimensions $(\ell, b, v) = (721, 361, 1024)$ representing the entire sky. The observed sky at $\delta > -30°$ fills about 70% of the data cube. The H I data discussed here form a subset of the entire H I data cube, within the limits $160° \leq \ell \leq 240°$, $-60° \leq b \leq 0°$. These limits encompass the Orion-Eridanus region as well as the Taurus and Perseus constellations.

## 3. H I loops, filaments and shell fragments in Orion-Eridanus

The Orion-Eridanus Bubble was identified earlier in H I by Heiles (1976) and Heiles & Jenkins (1976). Much of the H I structure that we discuss below is also visible in maps presented by Heiles (1984). Figure 1 shows a collection of velocity slices through our H I subset, each of width one $\rm km\,s^{-1}$, between $-40 \leq v_{\rm LSR} \leq +40\,{\rm km\,s^{-1}}$. The central velocity is indicated in each frame. The various structures must be traced in both spatial- and velocity-space, as is generally the case in H I analyses.

Figures 2 through 6 show individually-scaled maps of the column-density measure $\sum T_{\rm b} \Delta v (\rm K\,km\,s^{-1})$ for the indicated velocity intervals. These intervals were chosen after visually identifying filaments and partial loops in Fig. 1. Adjacent to the more-conventionally displayed H I data, Figs. 2–6 also show an unsharp-masked version of the maps. In this process (originally applied to photographic plates to enhance the dynamic range of low-amplitude fluctuations on high-intensity backgrounds by Malin, 1979), we subtracted a $10° \times 10°$ boxcar-smoothed copy of the image from the original. The size of the smoothing kernel roughly corresponds to the size of the intensity variations of the large-scale structures. The resulting maps bring out the filamentary structures beautifully. The filaments and loops that we identify as part of the expanding shell are shown isolated in Fig. 8. The maps in this figure are bitmaps showing the features for the same velocity intervals as maps A–E.

For orientation purposes Fig. 7 shows the location of the Orion OB1 association with respect to the H I shell. The dots in this figure show the positions of the brightest stars in the Orion constellation and the contours show the outlines of the Orion A and B molecular clouds as observed at 100 $\mu$m (IRAS). The ring around $(\ell, b) = (195°, -12°)$ is the $\lambda$-Orionis ring (see e.g., Wade 1957). The locations of the three main subgroups of Orion OB1 are indicated by circles.

*Map A* $(-40, -30)\,\rm km\,s^{-1}$: The emission in this extreme velocity interval is generally weak in regions away from the Galactic plane, but in the unsharp-masked map one can nevertheless

---

[1] The Dwingeloo 25 m telescope is operated by the Netherlands Foundation for Research in Astronomy (NFRA).

trace a number of filaments forming a ring-like structure. The filaments are located between $b = -50°$ and $-30°$ and $\ell = 175°$ and $210°$. These structures correspond to the approaching side of an expanding H I shell. At less extreme velocities, they merge with the features in Map B. The structures seen in this map can also be followed in Fig. 1, down to $v_{LSR} \approx -40$ km s$^{-1}$.

<u>Map B $(-28, -3)$ km s$^{-1}$</u>: A gap in the H I at negative velocities is clearly visible between latitudes $-50°$ and $-20°$ and longitudes $170°$ and $210°$. The filament at $\ell = 210°$ was used by BSNGG to derive the mass of the expanding H I shell. The H I gap has an irregular structure due to emission at $(\ell, b) = (190°, -30°)$. We suggest in Sect. 3 that this emission is probably associated with foreground material.

<u>Map C $(-1, +8)$ km s$^{-1}$</u>: The largest coherent H I structure in the Orion-Eridanus region can be seen in this map. It is the closed loop running between $\ell = 170°$ and $230°$ and $b = -55°$ and $-17°$. This structure is outlined well in the unsharp-masked image, with the filament at $b = 50°$ especially prominent. The cavity in H I surrounded by the loop structure shows emission which can be attributed (partly) to foreground material (see Sect. 3). The angular dimension of the loop perpendicular to the Galactic plane is about $33°$ and along the Galactic plane its maximum dimension is about $38°$. If we regard the loop as roughly circular, then its radius is about $18°$. If this H I shell is at the distance of the Orion OB1 association derived in Paper I as 380 pc, this implies a linear radius of about 120 pc. Also visible in this map is H I associated with the molecular cloud complex near the Orion OB1 association. The Orion A and B clouds are located near $(\ell, b) = (208°, -16°)$. The $\lambda$-Orionis ring is visible surrounding $(\ell, b) = (195°, -12°)$.

<u>Map D $(+9, +14)$ km s$^{-1}$</u>: The most prominent feature at the velocities in this map is the filament located around $b = -50°$. It is a continuation of the loop structure identified in Map C. The loop is now seen running between $\ell = 170°$ and $220°$ and $b = -55°$ and $-20°$. It has decreased in size at these velocities compared to those entering Map C in accordance with the expectations for an expanding shell.

<u>Map E $(+15, +40)$ km s$^{-1}$</u>: The most noticeable feature in this map is the almost-closed elliptical structure located between $\ell = 170°$ and $215°$ and latitudes $b = -53°$ and $-30°$. This ellipse can be followed in Fig. 1 up to $v_{LSR} \approx +40$ km s$^{-1}$, and represents the receding side of the H I shell.

From detailed inspections of these maps we conclude that H I features associated with the Orion-Eridanus Bubble can be followed in velocity space from $v_{LSR} = -40$ km s$^{-1}$ to $v_{LSR} = +40$ km s$^{-1}$.

## 4. Data at other wavelengths

In this section we compare the H I data with published data at other wavelengths.

### 4.1. H$\alpha$ data

An extensive study of the H$\alpha$ emission in the Orion-Eridanus area was made by RO. They studied the emission line profiles and constructed a map of diffuse H$\alpha$ emission which shows two prominent H$\alpha$ features, namely Barnard's Loop and a filament running along right ascension $4^h$ (from $(\ell, b) = (181°, -31°)$ to $(\ell, b) = (196°, -40°)$). Their map reveals that the diffuse Galactic H$\alpha$ background intensity is enhanced in the area of the H I cavity. They furthermore derived an expansion velocity of about 15 km s$^{-1}$ for the ionized gas, based on the observed line splitting in the emission profiles. The largest approaching and receding velocities they observed were $-30$ and $+26$ km s$^{-1}$, respectively. They derived a temperature of $8000 \pm 2000$ K for the ionized gas inside the cavity. The ultraviolet radiation from the O stars in Orion OB1 was regarded as the source of ionization. As a model for the origin of the Bubble they proposed the occurrence of a supernova, or series of supernovae, about $2 \times 10^6$ years ago, with a total energy of about $3 \times 10^{52}$ ergs. The Orion OB1 association is itself the most probable location for the suggested supernovae.

### 4.2. X-ray data

The cavity in the neutral hydrogen distribution in Orion-Eridanus also coincides with an enhancement in the background X-ray intensity. The most comprehensive study of the X-ray data in this region was carried out by BSNGG. They combined their X-ray data with neutral hydrogen data from the Bell Labs 21 cm survey (Stark et al. 1992) and with dust data from the IRAS mission. Maps were presented of the X-ray emission in two energy bands (0.25 keV and 0.6 keV). Two components in the morphology of the X-ray data were distinguished by BSNGG. There is a hook-shaped feature, corresponding to the morphology of the cavity in the H I emission apparent in Map D, which BSNGG referred to as EXE1 (located between $b = -50°$ and $b = -30°$ and $\ell = 180°$ and $\ell = 210°$), and a circular feature, corresponding to the H I cavity seen in Map C at $(\ell, b) = (210°, -43°)$, referred to as EXE2. The circular feature only shows up in the low-energy X-ray data. They derived temperatures of the X-ray emitting gas of $2.1 \times 10^6$ and $1.6 \times 10^6$ K for EXE1 and EXE2, respectively. Both X-ray features are associated with distinct H I shells, where EXE1 is associated with the shell that surrounds the Orion-Eridanus Bubble. The column density and mass of this shell were derived from the filament at $\ell = 210°$, visible in Map B, and from the approaching cap, at negative velocities. The mass derived from the approaching cap is an order of magnitude smaller than that derived from the filament.

A model for the Orion-Eridanus Bubble was proposed by BSNGG in which EXE1 is caused by a stellar-wind-blown cavity, and in which EXE2 is a separate bubble of hot gas possibly due to the wind from a single star (see their Fig. 9). In their model the Orion-Eridanus Bubble interacts with local clouds, roughly 100 pc from the Sun. Evidence for this interaction comes from the low-energy X-rays which appear to be partly absorbed by a filament of gas corresponding to the H$\alpha$ filament found by RO running along $\alpha = 4^{\mathrm{h}}$. Furthermore, a map of the ratio of X-ray emission in the high-energy band to the total emission shows evidence for absorption of low-energy X-rays by gas associated with the cloud Lynds Dark Nebula (LDN) 1569 at $(\ell, b) = (189°, -36°)$. To explain the discrepancy in the column densities derived from the filament at $\ell = 210°$ and the cap, BSNGG suggest that the Bubble is elongated along the line of sight and expands mostly towards the Sun (away from the Orion A and B clouds).

Figures 9a–d compare contour plots of the BSNGG X-ray data with gray-scale images of H I emission. The velocity intervals for which the H I data are shown were chosen to bring out the anti-correlation of the X-ray emission with kinematically-narrow features in the H I emission. The velocity intervals are $-1.0$ km s$^{-1}$ to $+4.0$ km s$^{-1}$ in Figs. 9a and 9b, and $+13.5$ km s$^{-1}$ to $+18.5$ km s$^{-1}$ in Figs. 9c and 9d. These velocity intervals correspond roughly to those in Map C and Map D. EXE2 was found by BSNGG to be associated with a separate H I shell surrounding the circular cavity in Map C. Indeed, Fig. 9a shows the occurrence of the low energy X-ray emission within the circular H I cavity in Map C. However, Figs. 9c and 9d show that the X-ray emission from both energy bands corresponds most closely to the gap in H I seen in Map D. This suggests that the X-ray-emitting gas is associated with a single coherent expanding H I shell. Based on the facts presented in Sect. 4.3, we argue that the appearance of a cavity in Map C around EXE2 may be caused by emission from foreground H I, partly filling in the H I cavity associated with the Bubble. If indeed the X-ray emission from both energy bands corresponds to a single cavity in H I, the nature of EXE2 remains to be investigated.

The greater detail available in the Leiden/Dwingeloo data (compared to the Bell Labs data utilized by BSNGG) reveals additional features in the H I morphology that anti-correlate with the X-ray emission. Figures 9c and 9d show the X-ray-absorbing H I gas associated with LDN 1569 at the lowest X-ray contour levels, which fits very well with the hook-shaped morphology of EXE1. Other lanes of absorbing H I gas can be seen in Figs. 9a and 9b, they run from $(\ell, b) = (207°, -33°)$ to $(\ell, b) = (199°, -38°)$ and from $(\ell, b) = (197°, -40°)$ to $(\ell, b) = (204°, -48°)$ (both lanes are located close to the $\alpha = 4^{\mathrm{h}}$ H$\alpha$ filament). The absorption by these H I lanes is visible in Fig. 9 as a decrease in the X-ray intensity, especially in the 0.25 keV energy band. This suggests that the absorption features found by BSNGG correspond to kinematically-narrow features in H I and more clearly establishes the association of the X-ray enhancement with the Orion-Eridanus Bubble.

If the X-ray-emitting gas is located inside an expanding H I bubble, one would expect the X-rays confined within it to be absorbed by the approaching, negative-velocity gas. The receding, positive-velocity gas would only attenuate X-rays emitted from the general background. For a spherical shell, the X-ray-emitting gas would be located in front of H I observed at positive velocities differing little from the expansion velocity. The anti-correlation between X-ray emission and H I emission indeed breaks down towards more positive velocities. This is shown in Fig. 10, where X-ray contours are overlaid on the gray-scale image of H I emission originating between $+20.0 \leq v_{\mathrm{LSR}} \leq +40.0$ km s$^{-1}$. The anti-correlation is still significant, suggesting an elongation of the H I shell along the line of sight.

Contrary to the conclusions of BSNGG, Fig. 9 shows that the X-ray enhancement is closely associated with the H I cavity at positive velocities. The H I features associated with the shell surrounding this cavity can be followed for velocities reaching up to about $+40$ km s$^{-1}$, as described in Sect. 3. From the bubble model proposed by BSNGG it follows that most of the H I emission from the Bubble should be visible at negative velocities. Our data show the contrary, suggesting that the Bubble is expanding more symmetrically with respect to the Orion A and B clouds.

### 4.3. CO data

The Orion-Eridanus region is too large to have yet been fully mapped in any molecular-line tracer. Most of the CO data available in this general area of the sky are confined to the neighborhood of the Orion Molecular Cloud complex. Two large molecular cloud complexes nearby are the Taurus and Perseus clouds, but these are located away from the H I shell and may be ignored in the present context. The OMC was most extensively covered in $^{12}$CO by Maddalena et al. (1986) and in $^{13}$CO by Bally et al. (1987). A comparison between H I and CO in the region of the Orion A and B clouds was made by Chromey et al. (1989). At high negative latitudes there is only limited coverage in the CO data, with only individual clouds studied. Such data can be found in Dame et al. (1987, down to $-25°$) and in Bally et al. (1991, down to $-30°$). A number of clouds detected in the high-latitude CO survey of Magnani et al. (1985, henceforth MBM) are located in the Orion-Eridanus area. The location of these clouds with respect to the H I can be seen in figures presented by Gir et al. (1994).

Several of the clouds mapped by MBM seem to have clear counterparts in our H I maps, namely MBM 110, 111, 16, 18, and 20. Gir et al. (1994) did detailed H I observations towards 10 regions containing 18 MBM clouds. They showed that in general the H I spectra of the high latitude clouds show several components with one component being clearly associated with the CO cloud. In particular Gir et al. confirm the association of MBM 16, 18 and 20 with H I. A statistical estimate of the distance to *all* the clouds in their sample was given by MBM as about 100 pc. Franco (1988) found a distance to MBM 18, specifically, of about 130 pc, based on $uvby\beta$ photometry. Magnani & De Vries (1986) found upper limits to the distances to MBM 18 and MBM 20 of 175 pc and 125 pc, respectively, based on star counts. The two clouds MBM 110 and MBM 111 are located just below Orion A at $(\ell, b) = (208°, -23°)$ and $(209°, -20°)$, respectively. Their velocities are about $+10$ km s$^{-1}$; a corresponding H I enhancement is seen in Map D. The cloud MBM 16 is located at $(\ell, b) = (172°, -38°)$, just at the edge of the H I shell visible in Map C. Its velocity is $+6.9$ km s$^{-1}$; an H I counterpart is visible as an enhancement in Map C. The clouds MBM 18 and MBM 20 are located inside the H I shell seen in Map C at $(\ell, b) = (189°, -36°)$ and $(\ell, b) = (211°, -37°)$ respectively. Their velocities are $+9.9$ km s$^{-1}$ and $+0.3$ km s$^{-1}$, respectively; H I counterparts are evident in Map C. (LSR velocities for MBM 16, 18 and 20 are from Gir et al. 1994.) Note that at the position of MBM 18 there is also enhanced H I emission in Map B. This indicates that H I emission seen at negative velocities in the direction of the Bubble may be due to foreground material. The cloud MBM 18 is associated with LDN 1569 and located close to the $\alpha = 4^h$ filament that appears to absorb the low energy X-rays; MBM 20 is associated with LDN 1642, in which a number of pre-main sequence objects is located (Sandell et al. 1987). Liljeström & Mattila (1988) mapped LDN 1642 in H I and found three velocity components at $-11, +0.5,$ and $+15$ km s$^{-1}$, plus a broad component extending from $-40$ to $+50$ km s$^{-1}$. This implies that the foreground material might be present over a large velocity interval in H I.

In their model BSNGG placed the clouds MBM 18 and 20 between the Local Bubble and the Orion-Eridanus Bubble and they suggested that these two clouds were formed as a consequence of the interaction of the Local Bubble and the Orion-Eridanus Bubble. This is consistent with the finding by Gir et al. (1994) that the H I gas associated with high latitude CO clouds is mostly in filamentary or loop-like structures. Furthermore Gir et al. suggest that the molecular clouds are formed in situ, possibly in the shock compressed gas in shells blown by stellar winds or supernovae.

There are two clouds visible in H I which have a cometary morphology. These are located at $(\ell, b) = (203°, -32°)$ and at $(206°, -26°)$, and can be seen in both the unsmoothed and filtered versions of the H I data shown in Map B. These structures are associated with the Lynds Bright Nebula (LBN) 917 and LBN 959, respectively, which were mapped in CO by Bally et al. (1991). The velocities are consistent with the H I counterpart velocities in Map B. From the cometary structure of these clouds, Bally et al. (1991) infer that they are actually located inside the Orion-Eridanus Bubble.

Thus we conclude that a substantial fraction of the H I emission found along the lines of sight toward the central cavity of the shell seen in Maps B and C is due to foreground material or to clouds located inside the Bubble.

### 4.4. IRAS data

Figure 11a shows a map of the 100 $\mu$m emission from dust in the Orion-Eridanus area. It is a mosaic of plates from the IRAS Sky Survey Atlas (Wheelock et al. 1994). (The sharp border seen running from $(\ell, b) \approx (220°, 0°)$ to $(\ell, b) \approx (195°, -30°)$ is due to a slight mismatch in the plate overlap.) A number of well known clouds, such as Orion A and B, and the $\lambda$-Orionis ring, can easily be identified in this map. The knot of emission near $(\ell, b) = (206°, -2°)$ is contributed by the Rosette Nebula; although the Rosette locally contaminates our H I emission maps over the range $+10$ to $+20$ km s$^{-1}$ (see Kuchar & Bania 1993), its distance of about 1600 pc places it well beyond the region of concern in this paper. There is also enhanced emission with respect to the surroundings at the positions of the clouds MBM 16, 18, and 20, as well as near LBN 917 and 959. The outlines of the H I shell visible in Map C are also seen in the 100 $\mu$m map, as well as a lane running from $(\ell, b) = (185°, -33°)$ to $(195°, -45°)$, which corresponds to the the H$\alpha$ filament mentioned in section 4.1.

Figure 11b shows a map of the total integrated H I emission in the Orion-Eridanus area. Note the tight gas-to-dust correlation revealed by the 100 $\mu$m dust image compared with the H I data integrated over the entire velocity range. (The cirrus/N(H I) correlation has been observed to be quite tight locally, and throughout the inner Galaxy generally; the correlation does seem to break down in the outer reaches of the Milky Way, where H I emission remains intense but the 100 $\mu$m emission becomes weak; see Burton 1992.) We note that the cirrus/N(H I) correlation which appears so tight in Fig. 11 can be kinematically unravelled, by comparing the dust emission with the H I contributed from narrow velocity slices of the sort shown in Fig. 1. Such unravelling has shown that cirrus features can sometimes be identified with H I emitting at quite anomalous velocities (see Deul & Burton 1990, and Burton et al. 1991). Establishing the velocity characteristic of various dust features in Fig. 11a confirms their association with the Orion-Eridanus region.

Figure 12 shows the correlation revealed by the H I gas column densities and the 100 $\mu$m dust intensities for different sections of the Orion-Eridanus area. The region that approximately encloses the H I shell can be delineated by the rectangle stretching from $170° \leq \ell \leq 230°$ and from $-55° \leq b \leq -30°$. Figures 12a and 12b show that the gas-to-dust correlation is not strong outside the region but that it is particularly tight within it. The latitude strips up to $b = -30°$ (Figs. 12c and 12d) show a similar behavior.

In studies confined near the Galactic equator, the correlation between N(H I) and $I_{100\mu m}$ in some cases yields a straight line where N(H I) gradually flattens off towards higher $I_{100\mu m}$ levels (see Deul 1988; Burton 1992). This flattening of the correlation can be explained as due to significant H I optical depths or to the variation of the interstellar radiation field. In Fig. 12e, there is such a pronounced flattening which we attribute to increasing optical depth of the H I. Figures 12a and 12f explicitly include regions lying quite close to the Galactic equator. Thus, although for latitudes closer to the equator than $b = -15°$ the intense H I emission may not be optically thin, Fig. 12a does show that the flattening originates from areas outside the H I shell. This implies that the H I emission from the Orion-Eridanus filaments and the cavity is optically thin, justifying our determination of the mass of the shell by directly converting the observed H I emission into column densities. The gas column depths at a high level are evidently contributed by gas lying along long lines of sight, transsecting the Galaxy at large, and therefore no doubt refer to gas not relevant to our discussion of the Orion-Eridanus Bubble.

## 5. Mass, energetics, and origin of the Orion-Eridanus Bubble

### 5.1. Mass of the Bubble

In order to determine the total mass of the H I in the shell surrounding the Orion-Eridanus Bubble we proceeded as follows. We isolated all the filaments and loops that we identify as part of the shell (see section 3) by marking those positions on the sky that coincide with the identified features. The filaments and loops that we identify as part of the expanding shell are shown in Fig. 8. In all positions coincident with the filaments the column densities were determined from the observed H I-emission by multiplying $\sum T_b \Delta v$ by $1.8224 \times 10^{18}$ (K km/s)$^{-1}$cm$^{-2}$. The observed column density at each position on the sky can then be converted into an H I mass.

Prior to the determination of the column density we subtracted a smooth background from each map to take into account the H I emission from regions along the line of sight that are not associated with the shell. The background was determined by fitting bivariate polynomials to the H I maps or by simply determining a constant background level. The total mass one finds differs depending on the background subtraction (the degree of the fitting polynomial). By determining the mass using the different background subtractions one gets a handle on the uncertainty in the mass estimate. The mass we determined is $2.3 \pm 0.7 \times 10^5 d_{380}^2$ M$_\odot$, where $d_{380}$ is the distance to the Orion-Eridanus Bubble in units of 380 pc. Our mass estimate includes a factor of 1.4 to account for the total-mass to H I-mass ratio.

Further causes of uncertainty in the mass estimate of the shell are: possible misidentification of some features in the H I maps, errors made in isolating the features, and the distance to the features (which were now all placed at 380 pc).

The total H I-mass estimates of the Bubble made by other authors are $4.7 \times 10^5 d_{380}^2$ M$_\odot$ (Heiles 1976), $2.6 \times 10^4 d_{380}^2$ M$_\odot$ (BSNGG), and $4.3 \times 10^3 d_{380}^2$ M$_\odot$ (BSNGG, using the negative-velocity cap). We note that the masses found by BSNGG are considerably smaller, due to the smaller size they infer for the shell surrounding EXE1.

### 5.2. Energy and origin of the Bubble

Following the discussion in Sect. 5.2 of Paper I, we investigate whether the Bubble could have been formed by the stellar winds and supernovae from Orion OB1. In the simplest approximation we assume the shell to be spherical and in that case the observed velocity extent of the shell corresponds to its expansion velocity. From our estimate of its mass, $2.3 \times 10^5$ M$_\odot$, and its expansion velocity, 40 km s$^{-1}$, we can calculate the kinetic energy of the expanding H I shell. If we assume that the bubble is expanding uniformly at this velocity then the kinetic energy is $3.7 \times 10^{51}$ ergs. We followed the model of wind-driven bubbles given by Castor et al. (1975) and Weaver et al. (1977), assuming a spherical shell embedded in a homogeneous ambient medium. The relevant equations for the shell radius, $R_s$, and expansion velocity, $V_s$, are those given by Shull (1993) (see Paper I).

This model predicts that some 20% of the total power released into the ISM by an OB association is converted into kinetic energy of the shell surrounding the wind-blown bubble. This implies that the Orion-Eridanus Bubble requires a total energy output over the lifetime of Orion OB1 of about $1.9 \times 10^{52}$ ergs. According to Paper I the total energy output from Orion OB1 is approximately $10^{52}$ ergs. This number includes contributions from both supernovae and stellar winds. We note that in modelling the evolution of the shell the mechanical luminosity from the association is considered constant in time. Thus the supernova contribution is averaged over the lifetime of the association and no distinction between a wind-driven or supernova-driven bubble is made. However, both contributions are important, with the stellar winds contributing 25% of the total energy output in the case of Orion OB1 (see Paper I). The total energy output from Orion OB1 is less than the energy required here by a factor of 2. The discrepancy may be due to the uncertainties in the estimated mass of the shell. It seems at least plausible from the energy requirements that the Orion-Eridanus Bubble has been blown into the ISM by the stellar winds and supernovae in Orion OB1.

From the estimated mass of the Bubble and its observed size we derive the initial ambient density of the ISM to be about 0.9 cm$^{-3}$. This is a large value at such high $|z|$. The H I volume-density value of 0.5 cm$^{-3}$ found generally characteristic close to the Galactic equator does, however, represent averaging over Galaxy-scale volumes of space. Such a general value may not be valid for the Orion-Eridanus region which is known to be one of intense star formation and therefore will be a site of generally high densities. We note that the mass estimate is uncertain and, furthermore, that the assumption of a spherical geometry for the Bubble may have led to an underestimate of its volume. (If an apparently spherical shell is elongated along the line of sight its volume will be larger than in the spherical case.) As discussed in Paper I, we can calculate $R_s$ and $V_s$ using only the energy output from Subgroup 1a of Orion OB1. If we assume that the age of the Bubble equals the age of this subgroup (11.4 Myr, see Paper I), the energy output from Subgroup 1a is about $2 \times 10^{37}$ ergs s$^{-1}$. This yields a value for $R_s$ which is larger (210 pc) and a value for $V_s$ which is lower (11 km s$^{-1}$) than the values which we found above. For an initial ambient density of half the value of 0.9 cm$^{-3}$, these numbers are 240 pc and 13 km s$^{-1}$, respectively. If we demand that the age be consistent with our observed values of $R_s$ and $V_s$, it follows that the age of the Bubble is $\sim 1.8$ Myr and the corresponding mechanical luminosity is $\sim 4 \times 10^{38}$ ergs s$^{-1}$. This value for the age of the shell is significantly different from the age of the oldest subgroup in Orion OB1, and the required luminosity is too high.

We note, however, that Orion OB1 is not located in a homogeneous ambient medium and that the association is located off-center inside the bubble (see Fig. 7). The first step in refining the model for the bubble is to take into account the fact that on *average* the Galactic H I layer is stratified perpendicular to the plane of the Galaxy. In the simplest approach the Galactic H I layer is approximately plane parallel. We would like to stress here that this stratification holds only on average. In localized regions (such as the Orion-Eridanus region) in the Galaxy one can find large density enhancements or underdense regions. Density variations that existed before the stellar winds and supernovae occurred will affect the expansion of the bubble. Comparison of a model which only takes the plane parallel density stratification into account to the observations will provide insight into these initial density inhomogeneities.

No simple analytical results have been derived for H I bubbles expanding in plane parallel stratified media, but various numerical simulations have been performed (see Tomisaka & Ikeuchi 1986; Mac Low & McCray 1988; Tenorio-Tagle & Bodenheimer 1988). For the case of adiabatic bubbles expanding into spherically symmetric media containing a finite amount of mass (implying density stratification), a semi-analytical treatment has been given by Koo & McKee (1990). Their model can be generalized to the plane parallel case (see below). The effects of the plane-parallel geometry of the Galactic H I are that bubbles will grow much larger in the direction perpendicular to the plane and might experience blow out (see Mac Low & McCray 1988). If the association that causes the bubble is located above or below the Galactic plane then the bubble will grow larger in the direction away from the plane. This explains the off center location of Orion OB1 with respect to the bubble. Note that in a plane parallel medium the expansion of the shell will not take place at a constant velocity over its surface. This means that the observed extent in velocity space no longer corresponds to the expansion velocity but rather indicates the lower limit on the actual expansion velocities in the shell surrounding the bubble.

In finite (stratified) media the shell surrounding the bubble will eventually accelerate and possibly blow out from the Galactic disk. Throughout the evolution of the bubble the observed expansion velocities will be higher than expected on the basis of the theory for a uniform medium. The models by Koo & McKee (1990) allow the calculation of the evolution of the shell of swept up matter surrounding the expanding bubble, given the central density and the mechanical energy input. The calculations for the spherically-symmetric case can be generalized to plane-parallel media, by assuming that all mass elements in the shell move on radial tracks (sector approximation) away from the source of energy. The equations for the one-dimensional case are assumed to remain valid. This implies that along the $z$-axis the results are the same as for the one-dimensional case, and along the direction perpendicular to the $z$-axis the solution will be the same as for a uniform medium. The approximation used by Koo & McKee underestimates the sizes of the bubbles, but is accurate to within 10% prior to the acceleration of the shell, as compared to numerical calculations. Once the shell has accelerated, the approximation is accurate to within 20%.

The Orion-Eridanus Bubble is located close to the Sun and its size is comparable to its distance. This means that projection effects are very important when one is observing the Bubble. Thus, when modelling the Bubble one should construct a three-dimensional model for the shell surrounding the Bubble and project the shell onto the sky in order to compare the model to the observations. We do this by calculating models with the Koo & McKee formalism for the two-dimensional case and rotating the resulting model around the axis perpendicular to the Galactic plane. The expansion center of the model (the Orion OB1 association) is placed at 380 pc from the Sun and 124 pc below the Galactic plane at a Galactic longitude of 205°. The models are compared to the observations under the requirement that the high latitude projection of the model matches the shell structures at $b \approx -50°$.

The Galactic H I layer has been described as consisting of two Gaussian components, with FWHM of 212 and 530 pc, and a broad exponential component with a scale height of 403 pc. The densities in the Galactic plane of these components are 0.395, 0.107, and 0.064 cm$^{-3}$, respectively (Dickey & Lockman 1990). Calculation of the expected $z$-profile of the Galactic H I layer in a realistic potential was done by Malhotra (1994). She shows that the result can approximately be described by a Gaussian $z$-distribution with a scale height of 144 pc and a mid-plane density of 0.86 cm$^{-3}$. Both density distributions are used in the model calculations. The results are shown in Fig. 13. In both cases we show map C overlaid by the model. The model consists of a grid of points on the surface of the shell and is projected into an $(\ell, b, v)$ data cube with the same spatial and kinematical resolution as the H I data cube. The overlaid model points are all those points on the surface of the shell that can be observed in the velocity interval corresponding to map C. Note that we do not model the thickness of the swept up shell. The irregular structure in the model data is due to the finite resolution with which the surface of the shell is represented.

When we calculate the shell evolution for the Dickey & Lockman density distribution the best match to the $b \approx -50°$ filaments is for an age of 5.3 Myr. The velocity range spanned in the projected model is $-19$ to $+24$ km s$^{-1}$ and the actual maximum velocity is 29 km s$^{-1}$ (note that due to projection effects the model shell is not symmetrical in velocity space). The bottom of the shell is located at 330 pc below the Galactic plane according to this model. For the model in which the Gaussian density distribution is used the best match is at an age of 5.8 Myr. The velocity range spanned in the projected model is $-17$ to $+23$ km s$^{-1}$ and the actual maximum velocity is 29 km s$^{-1}$. The bottom of the shell is located at 335 pc below the Galactic plane.

In both cases the model shell is much too large in the direction of the Galactic plane, as well as towards large longitudes. This indicates that indeed the assumed density stratifications are not appropriate for the Orion-Eridanus area. The Orion OB1 association is located close to the Orion molecular clouds and they probably constrain the expansion of the shell. The expansion also seems to be halted by density enhancements located between the Orion and Taurus molecular clouds (around $(\ell, b) \approx (180°, -20°)$). And indeed there are large H I column densities in that region. The expansion seems to be halted also by density enhancements along the bright ridge that runs from $(\ell, b) = (170°, -40°)$ to $(\ell, b) = (180°, -24°)$ in Map C. The observed shell is, however, larger than the models if one looks at the lower longitude side. This may indicate the presence of an underdense region around $(\ell, b) \approx (185°, -40°)$, into which the shell expands more easily.

One could evolve the models further in time to have them match the loop structure at $(\ell, b) \approx (190°, -50°)$, but that would require a confinement of the expansion in the direction $(230°, -50°)$. There are no obvious density enhancements there.

When looking at higher velocities the loops associated with the shell should decrease in size and eventually form a closed cap. But if we compare our model calculations to the observations at higher velocities we see that the decrease in size is much more rapid for the model than for the observed shell. This is already indicated in the channel maps of figure 1, where the elliptic structure at positive velocity shows no significant change in size. This discrepancy suggest that the observed shell is elongated along the line of sight. This was already suggested by BSNGG (see also section 4.2).

The dynamical age of the Bubble is consistently lower than the age of the oldest subgroup in Orion OB1. This may be explained by considering that a young subgroup first had to disrupt its parental molecular cloud, through the combined action of ionizing radiation and stellar winds, before it could blow a bubble into the ambient medium. In that case, not all of the energy from Subgroup 1a would have gone into the Bubble, but the energy from Subgroups 1b and 1c will have contributed as well. Furthermore, we have assumed a constant mechanical luminosity for the Orion OB1 association. In reality the energy output is not constant and depends on the star formation history of the association. As shown by Shull & Saken (1994), in the case of coeval or continuous star formation a peak in the energy output will occur as massive stars evolve off the main sequence. The peak in the output can be 3–10 times the later energy deposition rate due to supernovae and Wolf-Rayet stars. This peak in the energy output rate will result in faster shell growth and higher expansion velocities than in models with constant luminosity. Furthermore, they show that the ages of bubbles are underestimated by constant luminosity models. The underestimate is especially severe in the case of non-coeval star formation, an example of which is the formation of subgroups in an association. These findings are consistent with the fact that the age of the shell according to our model is lower than the age of the association and that the observed expansion velocities are higher than the model velocities.

Another way in which mechanical energy from Orion OB1 may be deposited into the ISM is through runaway OB stars. There are three well known runaway stars associated with Orion OB1: $\mu$ Col (O9.5V), AE Aur (O9.5V) and 53 Ari (B2V) (Blaauw 1961, 1993). The space velocities of these stars relative to Orion OB1 are estimated to be 117, 140 and 40 km s$^{-1}$, respectively. The time elapsed since their ejection from the center of Orion OB1 would be 2.3, 2.4 and 7.3 Myr. This implies that early type runaways can travel distances comparable to the size of the Orion-Eridanus Bubble in short times. It is possible that there were more runaways in the past and that they have already exploded as supernovae. One supernova explosion (with an explosion energy of $10^{51}$ ergs, and located in the same plane parallel media as used above) in the center of the Orion-Eridanus Bubble (at $(\ell, b) \approx (195°, -35°)$) could cause a shell with the same size and velocity extent as the observed shell. However the swept up mass would be too small by an order of magnitude compared to the observed shell mass. Letting the supernova explode in a uniform medium of density $0.9$ cm$^{-3}$ (see above) would obviously result in the right swept up mass, but the expansion velocity of the shell would be too low. However, the occurrence of a supernova outside the shell due to a runaway could create a low density environment into which the shell expands more easily and at higher velocities. Alternatively, the runaway could explode far from the association in an already existent bubble and will then provide extra energy for the shell expansion from a position close to the edge of the shell. This could be the cause of observed high velocities, where parts of the shell have been re-accelerated by the supernova explosion.

We conclude that the size and expansion velocity of the Orion-Eridanus Bubble can be explained as the result of stellar winds and supernovae from Orion OB1.

## 6. Conclusions

We presented maps of neutral hydrogen emission from the area surrounding the Orion-Eridanus Bubble, and showed that the observed velocity extent of the Bubble is about $\pm 40$ km s$^{-1}$. The enhanced X-ray emission in this area of the sky anti-correlates in a detailed way with kinematically-narrow features in H I, clearly establishing the association of the X-ray enhancement with the Bubble. Based on published CO data we conclude that most of the emission observed in the direction where a cavity is expected in Maps B and C originates from foreground clouds. Evidently, the observed X-ray emission is associated with the inside of a single, coherent H I shell. From comparison of the IRAS 100 $\mu$m dust data with the H I data we concluded that the H I shell is optically thin, which allowed us to determine the mass of the shell. The mass of the shell surrounding the Bubble was found to be $2.3 \pm 0.7 \times 10^5$ M$_\odot$. Assuming spherical symmetry, the mass and the expansion velocity of the shell correspond to a kinetic energy to be $3.7 \times 10^{51}$ ergs. According to the standard model the energy output from Orion OB1 is just large enough to account for the kinetic energy of the Bubble. Using models for wind-blown bubbles which incorporate the density stratification of the Galactic H I layer, we showed that the Orion-Eridanus feature can be explained as a wind-blown bubble. However, density inhomogeneities in the ambient medium cause large discrepancies between our model and the observed shell. The energy source for this bubble is the mechanical luminosity from stellar winds and from supernovae in Orion OB1.

*Acknowledgements.* We thank D.N. Burrows for kindly providing us the BSNGG X-ray data used in this paper. We are grateful for discussions with Tim de Zeeuw, Eugene de Geus and Adriaan Blaauw. We thank the referee for comments that helped improve the final version of this paper. The Infrared

Astronomical Satellite (IRAS) was a joint project of NASA (U.S.), NIVR (The Netherlands), and SERC (U.K.). This research was supported in part by the Netherlands Foundation for Research in Astronomy (NFRA) with financial support from the Netherlands Organization for Scientific Research (NWO).

**Fig. 1.** Caption is included beneath Fig. 1

**Fig. 2.** [**Map A**] Integrated H I emission in the velocity interval $-40 \text{ km s}^{-1} \leq v_{\text{LSR}} \leq -30 \text{ km s}^{-1}$. Both a logarithmically scaled image (left) and an unsharp-masked representation (right) are shown. Filaments located between $b = -50°$ and $b = -30°$ correspond to the approaching side of the expanding H I bubble

**Fig. 3.** [**Map B**] Integrated H I emission in the velocity interval $-28 \text{ km s}^{-1} \leq v_{\text{LSR}} \leq -3 \text{ km s}^{-1}$. The logarithmically scaled image is shown on the left and the unsharp-masked representation on the right. The major portion of an H I shell is visible between $b = -50°$ and $b = -20°$. The emission inside this structure may well be largely contributed by foreground material

**Fig. 4.** [**Map C**] Integrated H I emission in the velocity interval $-1 \text{ km s}^{-1} \leq v_{\text{LSR}} \leq +8 \text{ km s}^{-1}$. The logarithmically scaled image is shown on the left and the unsharp-masked representation on the right. The largest coherent structure in Orion-Eridanus can be seen in these maps. The unsharp-masked image shows the loop between $\ell = 170°$ and $\ell = 230°$ and $b = -50°$ and $b = -17°$. The Orion clouds A and B are located near $(\ell, b) = (208°, -16°)$. The $\lambda$-Orionis ring is visible around $(\ell, b) = (195°, -12°)$ (see also Fig. 11)

**Fig. 5.** [**Map D**] Integrated H I emission in the velocity interval $+9 \text{ km s}^{-1} \leq v_{\text{LSR}} \leq +14 \text{ km s}^{-1}$. The logarithmically scaled image is shown on the left and the unsharp-masked representation on the right. The filament running between $\ell = 170°$ and $\ell = 220°$ and $b = -53°$ and $b = -20°$ has a large velocity extent. It can also be associated with the loop structure in Map C

**Fig. 6.** [**Map E**] Integrated H I emission in the velocity interval $+15 \text{ km s}^{-1} \leq v_{\text{LSR}} \leq +40 \text{ km s}^{-1}$. The logarithmically scaled image is shown on the left and the unsharp-masked representation on the right. The elliptical structure between $\ell = 170°$ and $\ell = 215°$ and $b = -53°$ and $b = -30°$ can be traced in Fig. 1 up to $v_{\text{LSR}} = +40 \text{ km s}^{-1}$; it represents the receding side of the Orion-Eridanus H I bubble

**Fig. 7.** The position of the Orion OB1 association with respect to the H I shell. The grey scale image is a logarithmically scaled representation of integrated H I emission in the velocity interval $-1 \text{ km s}^{-1} \leq v_{\text{LSR}} \leq +8 \text{ km s}^{-1}$. The contours outline the 100 $\mu$m (IRAS) emmission from the Orion A and B molecular clouds (the ring around $(\ell, b) = (195°, -12°)$ is the $\lambda$-Orionis ring). The dots show the brightest stars in the Orion constellation. The circles show the positions of the three main subgroups of Orion OB1. From right to left are shown 1a, 1b and 1c

**Fig. 8.** Bitmaps, corresponding to Figs. 2–5, that show the features that are identified as part of the expanding H I shell in **(a)** map A, **(b)** map B, **(c)** Map C, **(d)** map D

**Fig. 8e.** Bitmap, corresponding to Fig. 6, that shows the features that are identified as part of the expanding H I shell in map E.

**Fig. 9.** Anti-correlation of H I emission (gray-scale image) and X-ray (contours) emission. The H I data are logarithmically scaled, and the contour values for the X-ray data are in counts s$^{-1}$. **(a)** H I emission between $-1 \leq v_{\text{LSR}} \leq +4 \text{ km s}^{-1}$ and 0.25 keV (L1 band) X-ray emission. **(b)** H I as in (a) and 0.6 keV (M1 band) X-ray emission. **(c)** H I emission between $+13.5 \leq v_{\text{LSR}} \leq +18.5 \text{ km s}^{-1}$ and 0.25 keV X-ray emission. **(d)** H I as in (c) and 0.6 keV X-ray emission

**Fig. 10.** Anti-correlation between X-ray emission (contours, in counts s$^{-1}$) and H I emission (gray-scale image, logarithmically scaled) for the velocity interval $+20 \leq v_{\text{LSR}} \leq +40 \text{ km s}^{-1}$. Both the 0.25 keV (L1 band) emission (a) and the 0.6 keV (M1 band) emission (b) display a prominent anti-correlation with the positive-velocity H I gas from the lower-$b$ region of the H I cavity; the H I gas lying closer to the Galactic equator may well be ambient foreground material, not associated with the Orion-Eridanus Bubble

**Fig. 11.** Maps of the extended Orion-Eridanus region showing in (a) the 100 $\mu$m IRAS dust emission and in (b) the total integrated H I emission. The generally-tight correlation is considered in the text in specific velocity- and spatial regions

**Fig. 12.** Scatter diagrams of N(H I) vs. $I_{100\mu m}$ for different regions in the Orion-Eridanus area, shown separately in (a) for the region outside the approximately rectangular boundary of the shell, and in (b) for the region inside the shell. Panels (c)–(f) show in 15° wide latitude strips that the linear correlation between N(H I) and 100 $\mu$m emission is strong at $b < -30°$, but breaks down nearer the Galactic plane

**Fig. 13.** Map C overlaid with the model for the expanding H I shell for: **(a)** an H I layer with a Gaussian density distribution (scale height 144 pc and mid-plane density 0.86 cm$^{-3}$) and **(b)** an H I layer with a density distribution as given by Dickey & Lockman (1990). In both cases we show all the points on the surface of the shell model that can be observed in the velocity interval corresponding to map C. The irregular structure of the models is due to finite resolution with with the shell surface is represented